\documentclass{elsart}

\usepackage{graphicx}
\usepackage[colorlinks,linkcolor=black]{hyperref}

\usepackage{amsmath,units,textcomp}
\usepackage{ragged2e}
\usepackage{endfloat}

\graphicspath{{./}{figs/}{figs/output/}}

\usepackage{lineno}

\linenumbers

\newcommand{\direction}[1]{\ensuremath{\left[#1\right]}}
\newcommand{\directions}[1]{\ensuremath{\left\langle#1\right\rangle}}

\newcommand{\gammaprime}{\ensuremath{\mathrm{\gamma^\prime}}}

\begin{document}

\begin{frontmatter}


\title{Best-Fit Ellipsoids of Atom-Probe Tomographic Data to Study Coalescence %
of $\gammaprime$ ($\mathrm{L1_2}$) Precipitates in Ni-Al-Cr}


\author[label1]{Richard A. Karnesky\corauthref{cor1}}
\ead{karnesky@northwestern.edu}
\ead[url]{http://arc.nucapt.northwestern.edu/}
\corauth[cor1]{Corresponding author.}
\author[label1]{Chantal K. Sudbrack}
\author[label1,label2]{David N. Seidman}
\address[label1]{Department of Materials Science and Engineering, Northwestern %
University, Evanston, IL 60208--3108, USA}
\address[label2]{Northwestern University Center for Atom-Probe Tomography%
\ (NUCAPT), Evanston, IL 60208--3108, USA}

\title{}

\author{}

\address{}

\begin{abstract} 
An algorithm is presented to fit precipitates in atom probe tomographic data
sets as equivalent ellipsoids.  Unlike previous techniques, which measure only
the radius of gyration, these ellipsoids retain the moments of inertia and
principle axes of the original precipitate, preserving crystallographic
orientational information.  The algorithm is applied to study interconnected
\gammaprime-precipitates ($\mathrm{L1_2}$) in the $\mathrm{\gamma}$-matrix (FCC)
of a Ni-Al-Cr alloy.  The precipitates are found to coagulate along
\directions{110}-type directions.
\end{abstract}

\end{frontmatter}
\newpage
%
One of the principle challenges that analyzing three-dimensional atom-probe
tomographic (APT) results poses is the amount of raw data that the instruments
are now able to collect~\cite{Kelly-2007,Seidman-2007}; we have collected
continuous data sets as large as $\mathrm{2.1\times10^8}$~properly ranged atoms
from a single specimen.  It is necessary to extract information about spatial
and compositional measures, such as precipitate size from these data sets.  A
common method to measure the size of precipitates is to calculate a single
radius of gyration (or, from this, the Guinier radius)~\cite{Miller-2000,%
Miller-2004}.  While this technique works well for equiaxed, spheroidal
precipitates~\cite{Miller-2003,Kolli-2007}, many alloy systems investigated by
APT have non-spherical features~\cite{Soisson-1996,Cadel-2002,Heinrich-2003,%
Wolde-Giorgis-2003,Dumont-2005,Isheim-2005,Erlach-2006,Pereloma-2006,%
Andersson-2007}.
Improperly reconstructed data sets, which have preferential evaporation or
local-magnification effects, may also exhibit non-spherical
features~\cite{Blavette-2001}.  Techniques that measure the angular
eccentricity of features can be used to quantify the accuracy of a
reconstruction. The radius of gyration technique does not, however, retain
three-dimensional information concerning precipitate orientation.

In this article, a more general alternative to the radius of gyration is
presented and applied.  Best-fit ellipsoids have equivalent centroids, moments
of inertia, and principle axes for arbitrarily shaped precipitates.  The
crystallographic orientations of the resulting ellipsoids are then used to study
the coagulation-coalescence coarsening mechanism in a Ni-Al-Cr alloy, which
occurs when the \gammaprime-precipitate number density is large
($>10^{24}\mathrm{~m^{-3}}$) and the edge-to-edge distance between adjacent
\gammaprime-precipitates is small ($<2\mathrm{~nm}$)~\cite{Sudbrack-2006,%
Mao-2007}.

In lattice kinetic Monte Carlo simulations, a coagulation-coalescence coarsening
mechanism is reported~\cite{Mao-2007}.  This mechanism is caused by
non-equilibrium overlapping diffusion fields, which originate from the
long-range vacancy-solute binding energies and a small mean edge-to-edge
interprecipitate distance.  The non-equilibrium concentration profiles observed
at the \gammaprime-precipitate/$\mathrm{\gamma}$-matrix interfaces lead to a
higher interfacial free energy than for fully equilibrated
\gammaprime-precipitates.  The excess free energy of the region of overlapping
concentration profiles (``diffuse neck'') can decrease by changing the
concentration thereof into a well-formed neck~\cite{Mao-2007}.  Phase-field
simulations find that the rate of \gammaprime-precipitate coalescence is
increased when the $\mathrm{\gamma/}\gammaprime$-interfacial width is increased
artificially, thereby increasing the overlapping diffusion
fields~\cite{Zhu-2004}.  While nanometer-sized coagulated
\gammaprime-precipitates might have been observed experimentally, past studies
only commented on whether precipitates appeared to be
non-equiaxed~\cite{Beddoe-1984,Schmuck-1996,Sudbrack-2004},
necked~\cite{Sudbrack-2005,Sudbrack-2007}, or chemically
ordered~\cite{Sudbrack-2006}.  They did not explore the crystallographic
orientation for precipitate coagulation.

A Ni-5.2 Al-14.2 Cr (at.\%) alloy was melted under an Ar atmosphere and chill
cast.  Its chemical composition was verified by inductively coupled plasma
spectroscopy.  The alloy was homogenized for 24 h at 1300\textcelsius{}, which
resulted in coarse grains (0.5--2 mm diameter).  After homogenization, the alloy
was annealed at 900\textcelsius{} ($\mathrm{\gamma}$-phase field) and water
quenched to ambient temperature.  The solutionized alloy was sectioned and aged
for 4 h at 600\textcelsius{} and then quenched.  This treatment leads to the
greatest percentage of \gammaprime-precipitates that are interconnected by
necks ($\mathrm{30\pm4\%}$)~\cite{Sudbrack-2006}.  The specimens were cut,
ground, and then electropolished into APT tips.  Three separate APT runs of
ca.~$\mathrm{3\times10^6}$~atoms were collected on a first-generation 3-D
APT~\cite{Blavette-1993,Cerezo-1998} at a specimen temperature of
$\mathrm{40.0\pm0.3}$~K, a pulse fraction of 19\%, and a pulse repetition rate
of 1.5 kHz.  The computer programs \textsc{ivas} (Imago Scientific Instruments)
and \textsc{adam}~\cite{Hellman-2002} were used to analyze APT data.  The
$\mathrm{\gamma/}\gammaprime$ interface is delineated using a 9~at.\%~Al
isoconcentration surface~\cite{Hellman-2003} and the atoms contained within the
\gammaprime-surface were exported and segmented into individual
\gammaprime-precipitates by a modified envelope algorithm~\cite{Miller-2000,%
Miller-2004,Hyde-2001,Marquis-2002}.

The \gammaprime-precipitates are divided into three classes:%
\ (i) single \gammaprime-precipitates (uncoagulated, without a concave neck)
that are not cut by the surface of the analyzed volume;%
\ (ii) two or more coalesced \gammaprime-precipitates that are interconnected by
a concave neck;%
\ (iii) \gammaprime-precipitates cut by the analysis volume boundary.
Class~(i) accounts for 42\% of \gammaprime-precipitates analyzed, class~(ii)
accounts for 28\% of the \gammaprime-precipitates analyzed, and only 12\% of
them are formations of more than two \gammaprime-precipitates (the largest of
which is made up of five distinguishable \gammaprime-precipitates).  The
best-fit ellipsoid method yields quantitative results for all three classes; 
particularly class~(ii), which is important for understanding the
coagulation-coalescence mechanism of \gammaprime-precipitate coarsening.

\begin{figure}[Htb]
  \begin{center}
    \includegraphics[width=0.5\textwidth]{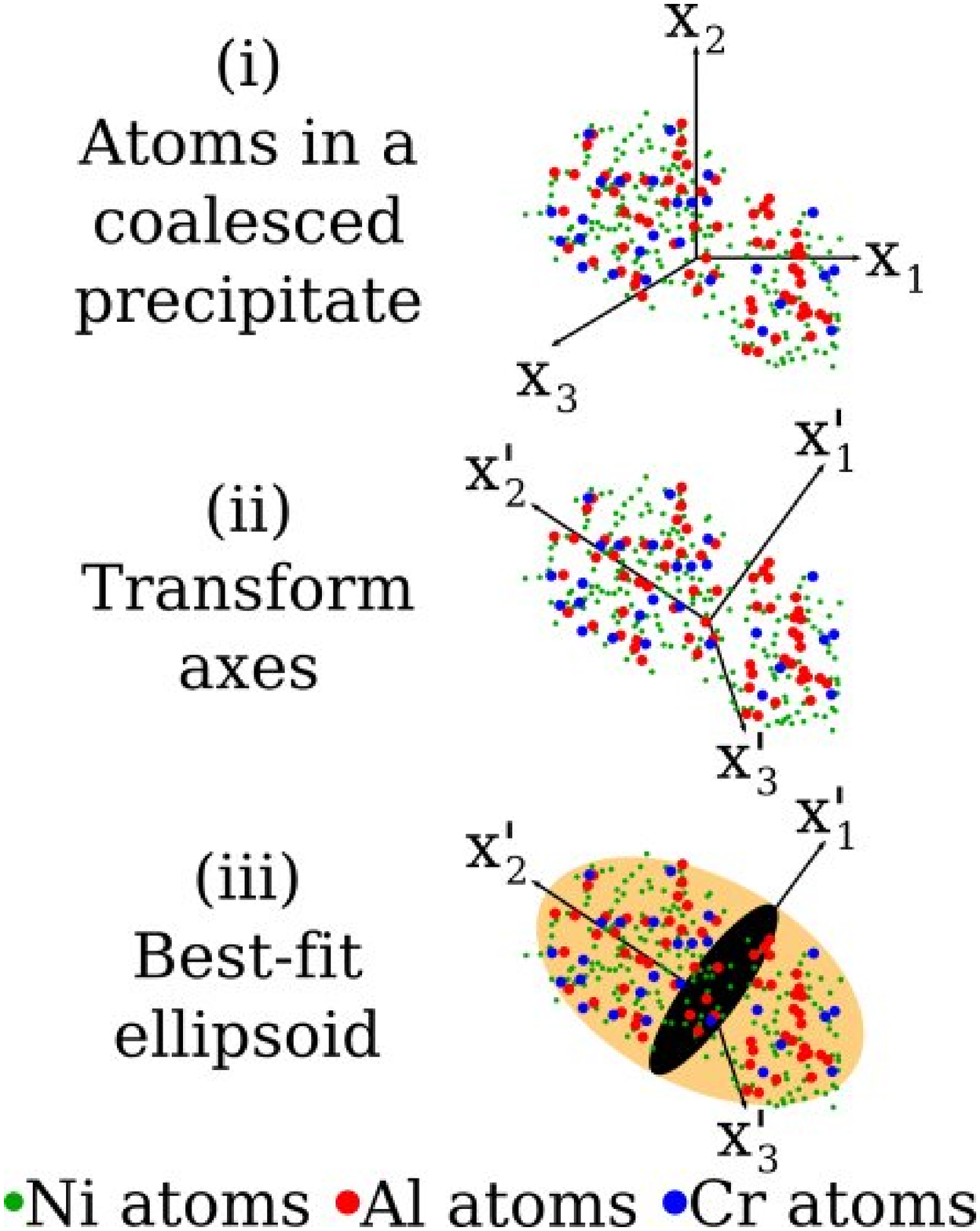}
  \end{center}
  \caption{The best-fit ellipsoid of a precipitate is determined in three steps:
  (i) for a reference set of axes ($X_1$, $X_2$, and $X_3$), identify the $x_1$,
  $x_2$, $x_3$ coordinates of all atoms in a precipitate and its center of mass
  to obtain a moment of inertia
  tensor ($L$) (Eqs.~\ref{eq:transform}--\ref{eq:average});
  (ii) the orientation of the principle axes ($X_1'$, $X_2'$, and $X_3'$) of the
  ellipsoid axes construct a Jacobian transformation matrix that will
  diagonalize $L$.
  (iii) the lengths of the major and minor semi-axes of the best-fit ellipsoid
  are found from the eigenvalues of the transformed matrix.
  (Eq.~\ref{eq:semi-axes}).}\label{fig:algorithm}
\end{figure}

A schematic that explains the fitting of an equivalent ellipsoid to atoms of
coagulated and coalesced precipitates is presented in Fig.~\ref{fig:algorithm}. 
For a reference space defined by Cartesian axes $X_1$, $X_2$, $X_3$ (typically
the analysis direction and the two orthogonal principle directions of the area
detector), the major and minor axes of the best-fit ellipsoid of a precipitate
containing $N$ atoms are determined directly from its eigenvalues ($\lambda_1$,
$\lambda_2$, $\lambda_3$), also referred to as the principle
axes~\cite{Nye-1985}.  The principle axes are segments along the transformed
Cartesian $X_1'$, $X_2'$, and $X_3'$ axes and are 
obtained from the diagonalization of the characteristic length matrix, $L$ (also
known as the inertia tensor).  This diagonalization is obtained by a Jacobian
transformation~\cite{Press-1986} of a symmetric second-rank tensor, as follows:

\begin{equation}
  L = \begin{bmatrix} l_{11} & l_{12} & l_{13} \\
                      l_{12} & l_{22} & l_{23} \\
                      l_{13} & l_{23} & l_{33} \end{bmatrix}
      \xrightarrow{\mathrm{transform}} 
      \begin{bmatrix} \lambda_1 & 0         & 0 \\
                      0         & \lambda_2 & 0 \\
                      0         & 0         & \lambda_3
      \end{bmatrix};\label{eq:transform}
\end{equation}

where the characteristic lengths, $l_{jk}$, are calculated from the positions of
$i^{\mathrm{th}}$ atom in the reference space ($x_1(i)$, $x_2(i)$, and
$x_3(i)$), relative to a precipitate's center of mass ($x_1(com)$, $x_2(com)$,
and, $x_3(com)$), averaged over $N$ atoms, employing:

\begin{align}
  l_{kk} &= \frac{1}{N}\sum_{i}^{N}\left(\sum_{j\neq{} k}%
    \left(x_j (i)-x_j (com)\right)^2\right) \\
  l_{jk} &= -\frac{1}{N}\sum_{i}^{N}\left(\left(x_j (i)-x_j (com)\right)%
                                     \left(x_k (i)-x_k (com)\right)\right)
                                     && \text{for $j\neq{} k$}\label{eq:average}
\end{align}

The diagonalization of the $L$ matrix follows a procedure outlined in
Ref.~\cite{Press-1986}.  The transformation matrix used for this diagonalization
yields the orientation of the ellipsoid with respect to the reference state.
Defining $\lambda_1\ge\lambda_2\ge\lambda_3$, the semi-axes ($S_i$) of the
best-fit ellipsoid are given by:

\begin{align}
  S_i &= \sqrt{\frac{5}{2}\left(\lambda_j+\lambda_k-\lambda_i\right)}
  && \text{for $j\neq k$};&&\label{eq:semi-axes}
\end{align}

where $S_3 \ge S_2 \ge S_1$ are the major semi-axis and two minor semi-axes,
respectively.

\begin{figure}[Htb]
  \begin{center}
    \includegraphics[width=0.5\textwidth]{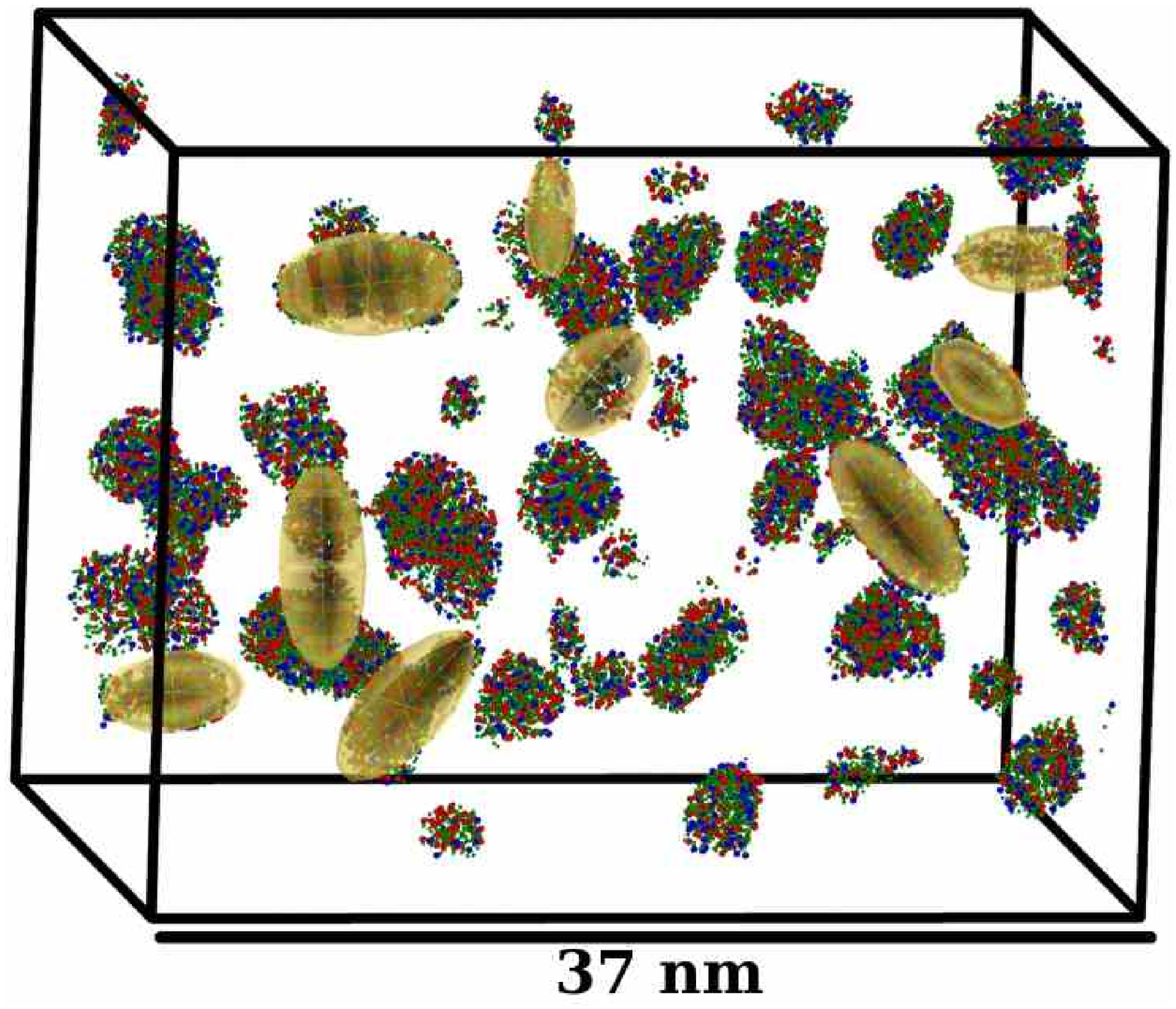}
  \end{center}
  \caption{A representative data set for a Ni-5.2 Al-14.2 Cr (at.\%) alloy,
  whose thermal history is discussed in the text.  The coloring scheme matches
  that of Fig.~\ref{fig:algorithm}, with Ni atoms in green, Al atoms in red, and
  Cr atoms in blue.  The atoms in the $\mathrm{\gamma}$-matrix (FCC) are omitted
  for clarity.  The gold colored best-fit ellipsoids indicate those
  \gammaprime ($\mathrm{L1_2}$) precipitates that are interconnected by
  necks.}\label{fig:apt}
\end{figure}

Figure~\ref{fig:apt} shows a reconstruction of one APT data set where ellipsoids
have been fit to interconnected \gammaprime-precipitates.  As is seen, the
best-fit ellipsoid retains 3-D size and crystallographic orientation information
that is jettisoned by other techniques~\cite{Miller-2000,Miller-2004}.

Single \gammaprime-precipitates that are not cut by the analysis volume have
aspect ratios of $\frac{S_3}{S_2}=1.5\pm0.5$ and $\frac{S_2}{S_1}=1.3\pm0.2$.
The closeness of these values to unity is consistent with equiaxed, uncoagulated
precipitates.

Coalesced \gammaprime-precipitates that are interconnected by necks have aspect
ratios of $\frac{S_3}{S_2}=2.9\pm0.9$ and $\frac{S_2}{S_1}=1.3\pm0.3$.  The
ratio for $\frac{S_3}{S_2}$ is about twice the same ratio for uncoagulated
precipitates, but $\frac{S_2}{S_1}$ is about the same for the two classes.  This
demonstrates that a majority of these consist of two equiaxed
\gammaprime-precipitates that have coagulated and undergone coalescence.

Precipitates that are cut by the edge of the analysis volume can serve as a
check of the best-fit ellipsoid method.  These have $\frac{S_3}{S_2}=2.2\pm0.8$
and $\frac{S_2}{S_1}=1.4\pm0.4$ because the majority are equiaxed, uncoagulated
\gammaprime-precipitates that are, on average, cut in two by the analysis
boundary.  Sampling bias by coagulated \gammaprime-precipitates is small, as
they make up a minority of precipitates and many can be isolated (if their
concave necks are in the analysis volume).

\begin{figure}[Htb]
  \begin{center}
    \includegraphics[width=0.5\textwidth]{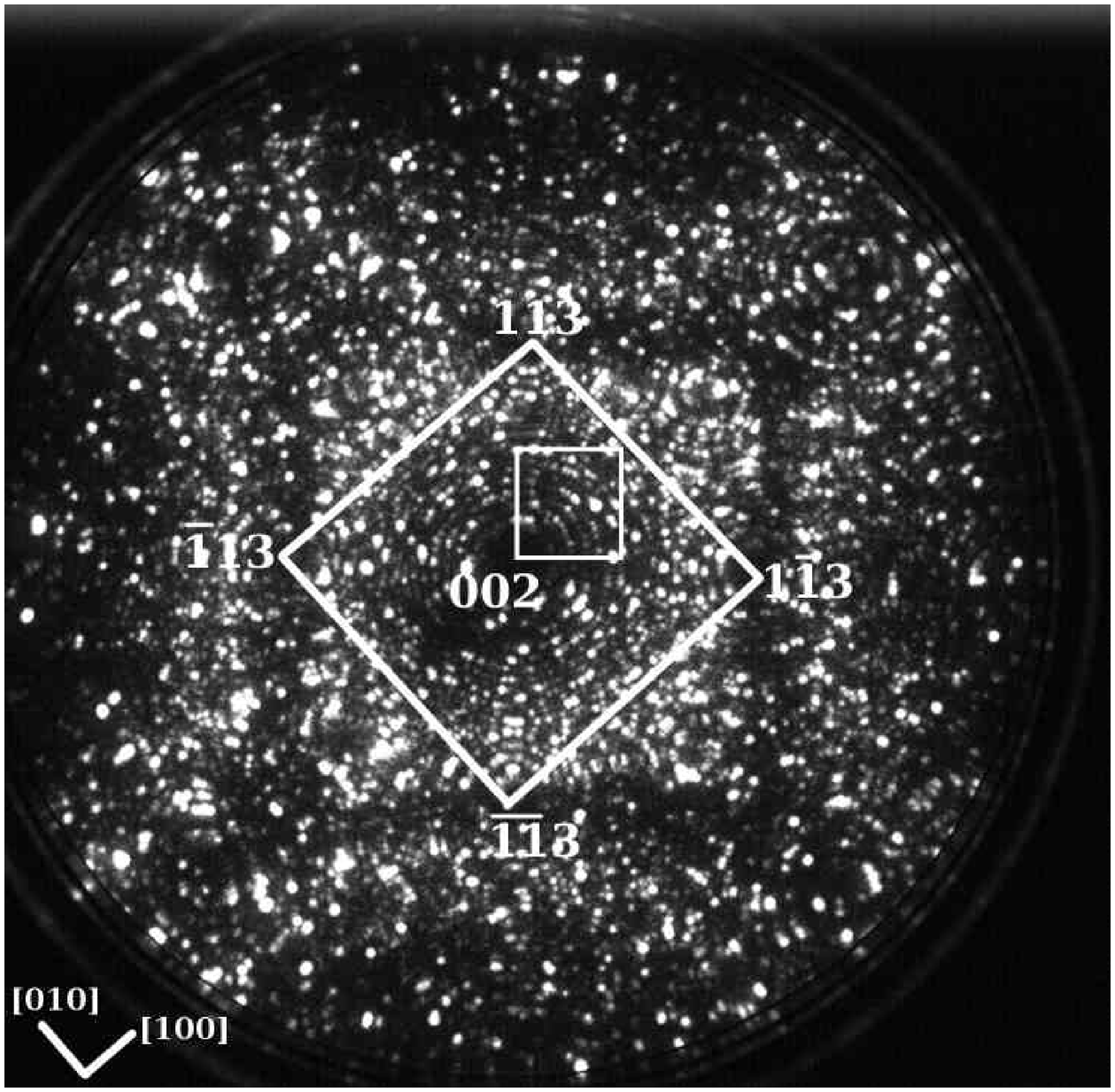}
  \end{center}
  \caption{An FIM image, centered on the 002 pole, taken before APT analysis.
  The small square denotes the area and orientation for the data set in
  Fig.~\ref{fig:algorithm}.  The larger square connects the 113 family of poles
  and the edges of this square give the \direction{010} and \direction{100}
  directions.}\label{fig:fim}
\end{figure}

It is useful to relate the axes of the analysis volume with specific
crystallographic directions to study the orientation of \gammaprime-precipitate
coagulation.  The analysis direction, which was chosen near the 002
crystallographic pole, provides us with the \direction{001} direction.  The
\direction{010} and \direction{100} directions can be deduced from a Field-Ion
Microscope (FIM) image (Fig.~\ref{fig:fim})~\cite{Bowkett-1970} to within
1--4\textdegree.  FIM micrographs also demonstrate that local magnification
effects are negligible in this alloy, which is consistent with the small lattice
parameter misfit, $\delta=0.0006\pm0.0004$~\cite{Sudbrack-2006}.

\begin{figure}[Htb]
  \begin{center}
    \includegraphics[width=0.5\textwidth]{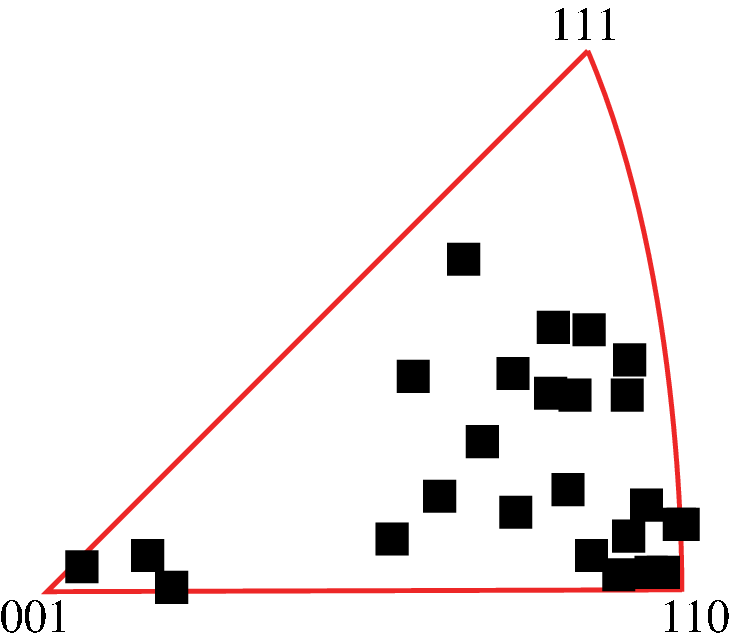}
  \end{center}
  \caption{An inverse pole figure for all interconnected \gammaprime{}
  ($\mathrm{L1_2}$) precipitates in this study, presented in the standard
  stereographic triangle.  There is a preference for coagulation along and close
  to \directions{110}-type directions, which is consistent with a diffusion
  controlled mechanism for coagulation.}\label{fig:ipf}
\end{figure}

The rotation matrix used in the best-fit ellipsoid method yields the Bunge Euler
angles~\cite{Bunge-1982} for the major principle axis of the
\gammaprime-precipitate with respect to this crystallographically-resolved
reference system.  From these, an inverse pole figure (Fig.~\ref{fig:ipf}) shows
the orientations of \gammaprime-precipitates that are the result of the
coagulation-coalescence coarsening.  There is a preference for coagulation
along the \directions{110}-type directions.  30\% of the coalesced precipitates
are within 10\textdegree{} and 71\% are within 15\textdegree{} of
\directions{110}-type directions.  In the FCC structure of the
$\mathrm{\gamma}$-matrix, \directions{110} is the fastest diffusion path for
solute clusters, and is therefore consistent with the model presented in
Ref.~\cite{Mao-2007}.  Some \gammaprime-precipitates coagulated along
\directions{100}, which is the next-fastest diffusion path and none coagulated
along \directions{111}, which is a slower path.

No crystallographic orientational preference was found for single, uncoagulated
\gammaprime-precipitates.  This supports, once more, the equiaxed nature of
\gammaprime-precipitates and the proper reconstruction approach (with negligible
local magnification) for analyzing the raw APT data.  The
\gammaprime-precipitates that are cut by the analysis boundary show a preference
for the \direction{100} direction, as that direction makes up the majority of
the analysis frustum's surface.

The measurement of size and orientation of non-spheroidal precipitates in APT
data requires more spatial information to be preserved than for the commonly
used methods currently reported to date.  We have demonstrated that the best-fit
ellipsoid technique preserves the center of gravity, moment of inertia, and
principle axes of any precipitate.  The technique is applied to specific results
for a Ni-Al-Cr alloy with both uncoagulated equiaxed \gammaprime-precipitates
and nonequiaxed lobed precipitates that formed through a coagulation-coalescence
coarsening mechanism.
\begin{ack}
This research is supported by the National Science Foundation, Division of
Materials Research, under contract DMR--0241928.  RAK received partial support
from a Walter P. Murphy (WPM) Fellowship and the US Department of Energy%
\ (DE--FG02--98ER45721).  CKS received partial support from an NSF graduate
research fellowship, a WPM Fellowship, and a Northwestern University terminal
year fellowship.  We thank Imago Scientific Instruments and Dr.\ M.\ K.\ Miller
for permitting RAK to modify the source code for \textsc{envelope}, and Dr.\ D.%
\ Mainprice for his inverse pole figure software, \textsc{ipf2k}.
Prof.\ D.\ C.\ Dunand, Dr.\ Z.\ Mao, and Dr.\ G.\ Martin are thanked for
discussions.
\end{ack}
{\RaggedRight{}
\bibliographystyle{elsart-num}
\bibliography{ellipsoid}}
\end{document}